\RequirePackage{fix-cm}
\documentclass[pdftex,twocolumn,epjc3]{svjour3}
\smartqed  % flush right qed marks, e.g. at end of proof
\RequirePackage{graphicx}
\RequirePackage[colorlinks,citecolor=blue,urlcolor=blue,linkcolor=blue]{hyperref} 
\RequirePackage{lineno}
\usepackage{color}
\usepackage{multirow}
\usepackage{booktabs} %better hlines in tables
\journalname{Eur. Phys. J. C}
\usepackage{amssymb}     % packages called in main file, e.g. gstr-template.tex
\usepackage{marvosym}    % symbols for EURO
\usepackage{upgreek}     % for upright greek letters in units
\newcommand{\mum}         {{$\upmu$m}}
\newcommand{\mus}         {{$\upmu$s}}

\newcommand{\baseT}[2]{\mbox{$#1{\cdot}10^{#2}$}}

\newcommand{\ga}          {$\gamma$}

\newcommand{\gline}       {$\gamma$ line}

\newcommand{\grays}       {$\gamma$ rays}

\newcommand{\onbb}        {{$0\nu\beta\beta$}}

\newcommand{\onecec}      {{$0\nu\rm{ECEC}$}}
\newcommand{\nnecec}      {{$2\nu\rm{ECEC}$}}
\newcommand{\etal}        {\textit{et al.}}

\newcommand{\gerda}       {\textsc{Gerda}}

\newcommand{\GERDA}       {\mbox{\textsc{Gerda}}}  

\newcommand{\LNGS}        {{\mbox{\textsc{Lngs}}}}

\newcommand{\lar}         {LAr}
\newcommand{\geni}        {{\mbox{\textsc{Genius}}}}

\newcommand{\bege}        {{\sc BEGe}}
\newcommand{\phaseone}    {Phase~I}

\newcommand{\IGEX}        {{\mbox{\textsc{Igex}}}}

\newcommand{\HDM}         {\mbox{\textsc{HdM}}}

\newcommand{\geant}       {\textsc{Geant4}}
\newcommand{\GEANT}       {\textsc{\mbox{{Geant}}}}

\newcommand{\mage}        {\textsc{MaGe}}
\newcommand{\gelatio}     {\textsc{Gelatio}}

\newcommand{\geenr}       {{$^{\rm enr}$Ge}}          %$^{\rm enr}$Ge
\newcommand{\genat}       {{$^{\rm nat}$Ge}}

\newcommand{\nuc}[2]      {{$^{#2}$\rm #1}}

\begin{document}

\title{Limit on the Radiative Neutrinoless Double Electron Capture of
  $^{36}$Ar from \mbox{\protect{\textsc{GERDA}}}  Phase~I}


\author{The \protect{\mbox{\textsc{GERDA}}} collaboration\thanksref{corrauthor}
        \and  \\[4mm]
 M.~Agostini\thanksref{ALNGS} \and
 M.~Allardt\thanksref{DD} \and
 A.M.~Bakalyarov\thanksref{KU} \and
 M.~Balata\thanksref{ALNGS} \and
 I.~Barabanov\thanksref{INR} \and
 N.~Barros\thanksref{DD,nowPEN} \and
 L.~Baudis\thanksref{UZH} \and
 C.~Bauer\thanksref{HD} \and
 E.~Bellotti\thanksref{MIBF,MIBINFN} \and
 S.~Belogurov\thanksref{ITEP,INR}
% ...
}

\thankstext{corrauthor}{INFN Laboratori Nazionali del Gran Sasso, Italy.\\
\emph{Correspondence},
                                email: gerda-eb@mpi-hd.mpg.de}
 
\thankstext{nowPEN}{\emph{present address:} Dept. of Physics and Astronomy,
  U. of Pennsylvania, Philadelphia, Pennsylvania, USA}
\thankstext{alsoMIPT}{\emph{also at:} Moscow Inst. of Physics and Technology,
  Russia} 
\thankstext{alsoIUN}{\emph{also at:} Int. Univ. for Nature, Society and
    Man ``Dubna'', Dubna, Russia}  
\thankstext{nowParis}{\emph{present address:}
LAL, CNRS/IN2P3,
 Universit{\'e} Paris-Saclay, Orsay, France}

\institute{
INFN Laboratori Nazionali del Gran Sasso and Gran Sasso Science Institute, Assergi, Italy\label{ALNGS} \and
INFN Laboratori Nazionali del Sud, Catania, Italy\label{CAT} \and
Institute of Physics, Jagiellonian University, Cracow, Poland\label{CR} \and
Institut f{\"u}r Kern- und Teilchenphysik, Technische Universit{\"a}t Dresden, Dresden, Germany\label{DD} \and
Joint Institute for Nuclear Research, Dubna, Russia\label{JINR} \and
Institute for Reference Materials and Measurements, Geel, Belgium\label{GEEL} \and
Max-Planck-Institut f{\"u}r Kernphysik, Heidelberg, Germany\label{HD} \and
Dipartimento di Fisica, Universit{\`a} Milano Bicocca, Milan, Italy\label{MIBF} \and
INFN Milano Bicocca, Milan, Italy\label{MIBINFN} \and
Dipartimento di Fisica, Universit{\`a} degli Studi di Milano e INFN Milano, Milan, Italy\label{MILUINFN} \and
Institute for Nuclear Research of the Russian Academy of Sciences, Moscow, Russia\label{INR} \and
Institute for Theoretical and Experimental Physics, Moscow, Russia\label{ITEP} \and
National Research Centre ``Kurchatov Institute'', Moscow, Russia\label{KU} \and
Max-Planck-Institut f{\"ur} Physik, Munich, Germany\label{MPIP} \and
Physik Department and Excellence Cluster Universe, Technische  Universit{\"a}t M{\"u}nchen, Germany\label{TUM} \and
Dipartimento di Fisica e Astronomia dell{`}Universit{\`a} di Padova, Padua, Italy\label{PDUNI} \and
INFN  Padova, Padua, Italy\label{PDINFN} \and
Physikalisches Institut, Eberhard Karls Universit{\"a}t T{\"u}bingen, T{\"u}bingen, Germany\label{TU} \and
Physik Institut der Universit{\"a}t Z{\"u}rich, Z{u}rich, Switzerland\label{UZH}}


\date{Received: date / Accepted: date}

\maketitle

\begin{abstract}
Neutrinoless double electron capture is a process that, if detected, would
give evidence of lepton number violation and the Majorana nature of neutrinos.
A search for neutrinoless double electron capture of $^{36}$Ar has been
performed with germanium detectors installed in liquid argon using data from
\phaseone\ of the GERmanium Detector Array (\GERDA) experiment at the Gran
Sasso Laboratory of INFN, Italy. No signal was observed and an experimental
lower limit on the half-life of the radiative neutrinoless double electron
capture of $^{36}$Ar was established: $T_{1/2} > $ 3.6 $\times$ 10$^{21}$ yr
at 90\,\% C.I.
\end{abstract}
%%%%%%%%%%%%%%%%%%%%%%%%%%%%%%%%%%%%%%%%%%%%%%%%%%%%%%%%%%%%%%%%%%%%%%%%%%%%%%%%

\section{Introduction}
\label{intro}
The observation of neutrinoless double beta decay (\onbb):
\begin{equation}
(A,Z-2) \rightarrow (A,Z) + 2e^-,
\end{equation}
can provide unambiguous information on lepton number violation and indicate
the Majorana nature of neutrinos, regardless the physics mechanism responsible
for the decay. Currently many experiments are searching for this decay
considering different isotopes.  Among these is the \GERDA~(GERmanium Detector
Array) experiment~\cite{gerdatec:2013} implementing bare germanium detectors
enriched in $^{76}$Ge. This experiment searches for neutrinoless double beta
decay of $^{76}$Ge.  Recently the best limit on \onbb\ decay half-life of
$^{76}$Ge has been published by the \GERDA\ collaboration~\cite{0nbbgerda:2013}.

Another lepton number violating process that can provide the same information
as neutrinoless double beta decay is the double capture of two bound atomic
electrons without the emission of neutrinos (\onecec):
\begin{equation}
2e^- + (A,Z+2) \rightarrow (A,Z) + Q,
\end{equation}
where the quantity $Q$ corresponds to the energy difference between
the ground state atoms $(A,Z+2)$ and $(A,Z)$~\cite{winter55,ber83}. 
\begin{figure}[b]
\begin{center}
  \includegraphics[width=.45\textwidth]{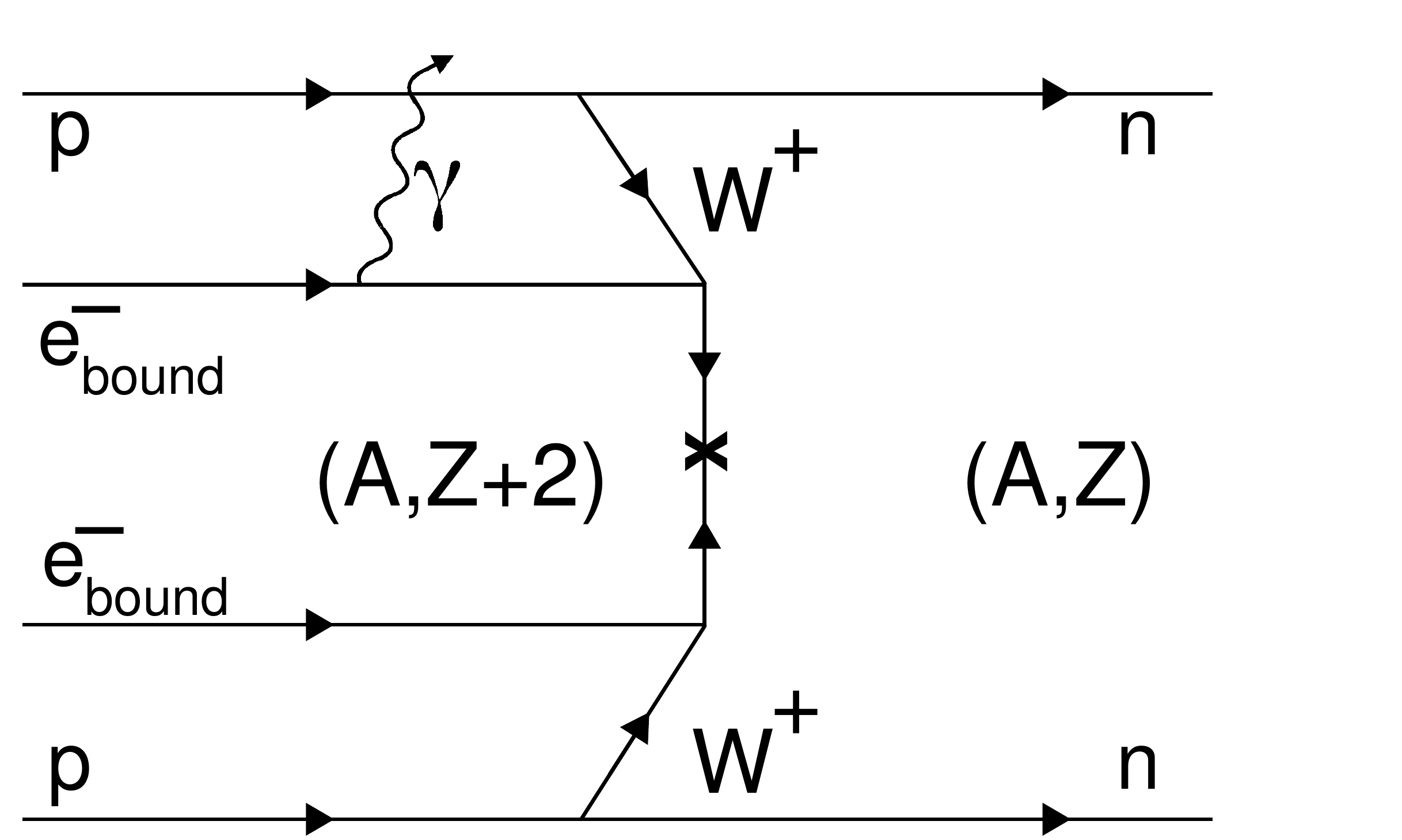}
 %figures/2ECdyagram.pdf}
\caption{\label{fig:0n2ECmode}       
         Diagram for zero neutrino double electron capture with the
         emission of one photon
}
\end{center}
\end{figure}
While in the corresponding process where two neutrinos are emitted (\nnecec)
the available energy of the decay is carried away by neutrinos plus X-rays or
Auger electrons, in the neutrinoless double electron capture the decay must be
accompanied by the emission of at least another particle to ensure energy and
momentum conservation. Different modes can be considered in which
\onecec\ decay is associated with the emission of different particles like
$e^+e^-$ pairs, one or two photons, or one internal conversion electron.  A
detailed discussion about double electron capture processes can be found in
Refs.~\cite{0nececreview:vergados,0nececreview:doikotani,0nececreview:simkovic}.

For $0^+\rightarrow0^+$ transitions the capture of two $K$-shell electrons
with the emission of only one photon is forbidden because of angular momentum
conservation. Therefore, the most likely process is the capture from the $K$-
and the $L$-shell.  The diagram of this mode is depicted in
Fig.~\ref{fig:0n2ECmode}. The unstable daughter atom relaxes by emission of
X-rays or Auger electrons.

At present, only two experiments found an indication of two neutrino double
electron capture. The first is based on a geochemical measurement of
$^{130}$Ba decay into $^{130}$Xe~\cite{meshik,pujol} and the second is a
large-volume copper proportional counter searching for double K-shell capture
in $^{78}$Kr~\cite{gavrilyuk}.  Several experiments including the latter
established limits on both neutrino accompanied and neutrinoless double
electron capture of different isotopes~(see
Refs.~\cite{gavrilyuk,tgvresult,mei,frekersresult,belli,povinec,barabash1,barabash2}).
For some isotopes the possibility of a resonant enhancement of the
\onecec\ decay has been predicted in case of mass degeneracy between
the initial state and an excited final state~\cite{ber83,eliseev}.

$^{36}$Ar is expected to undergo double electron capture to the ground state
of \nuc{S}{36}~\cite{tretyak}. The available energy of the decay is
432.6$\pm$0.2~keV and, therefore, both the radiative and the internal
conversion modes are energetically allowed~\cite{levelsref}.  A resonance
enhancement of the decay is not possible for this isotope.  Calculations based
on the quasiparticle random-phase approximation (QRPA) predict a half-life for
$^{36}$Ar in the order of 10$^{38}$~yr for an effective Majorana neutrino mass
of 1~eV~\cite{Merle:2009}.  So far, an experimental limit on the radiative
mode obtained during detector characterizations in the \GERDA\ Detector
Laboratory has been published ($T_{1/2}$ \baseT{>1.9}{18}~yr at 68\,\%
C.L.)~\cite{Chkvorets:2008wj}.

The radiative mode of \onecec\ in $^{36}$Ar with the emission of one
photon provides a clear signature through the discrete value of its energy and
allows the detector to be separate from the source of the decay.  Two cascades
of characteristic X-rays with energies of $E_K=2.47$ keV and $E_L=0.23$~keV
are emitted, corresponding to the capture of the electrons from the $K$- and
the $L$-shell, respectively.  The corresponding energy for the monochromatic
photon is $E_\gamma = Q - E_k - E_L = 429.88\pm0.19$~keV.

This paper reports the search for the 429.88~keV \ga\ line  from
\onecec\ decay of $^{36}$Ar with \GERDA\ \phaseone\ germanium detectors
and the determination of a limit on its half-life.

\section{The \GERDA\ experiment}
\label{sec:1}
The \GERDA\ experiment~\cite{gerdatec:2013} is located at the Laboratori
Nazionali del Gran Sasso (\LNGS) of the INFN.  It was designed in two phases.
During \phaseone\ reprocessed $p$-type semi-coaxial High-Purity Germanium
(HPGe) detectors enriched in $^{76}$Ge (\nuc{Ge}{\rm enr}) to up to
86\,\%~\cite{gerdadets} from the \HDM~\cite{hdm} and \IGEX~\cite{igex}
experiments have been employed in the experiment as well as natural germanium
(\nuc{Ge}{\rm nat}) HPGe detectors from the \geni\ Test Facility and newly
produced enriched Broad Energy Germanium (\bege) detectors~\cite{canberra1}.
The bare detectors are immersed into a cryostat containing 64~m$^3$ (89.2~t)
of \lar, which acts both as the coolant medium and a shield against external
radiation.  The isotopic abundance of $^{36}$Ar in natural argon is
0.3336(4)\,\%~\cite{Lee:2006bc}, which sums up to about 298~kg.  An additional
shield of ultra pure water (10~m in diameter) surrounds the cryostat
containing the argon. The water tank is instrumented with 66 PMTs as a muon
Cherenkov veto~\cite{muonveto}.  Each detector string is surrounded by a
60~\mum\ thick Cu foil (``mini-shroud''), to limit drifting of $^{42}$K ions
to detector surfaces. In addition, to mitigate radon contamination, a
30~\mum\ Cu cylinder (``radon shroud'') surrounds the array of strings.

\section{Data taking and data selection}
\label{sec:datataking}

\begin{figure*}[t]
\begin{center}
  \includegraphics[width=0.48\textwidth]{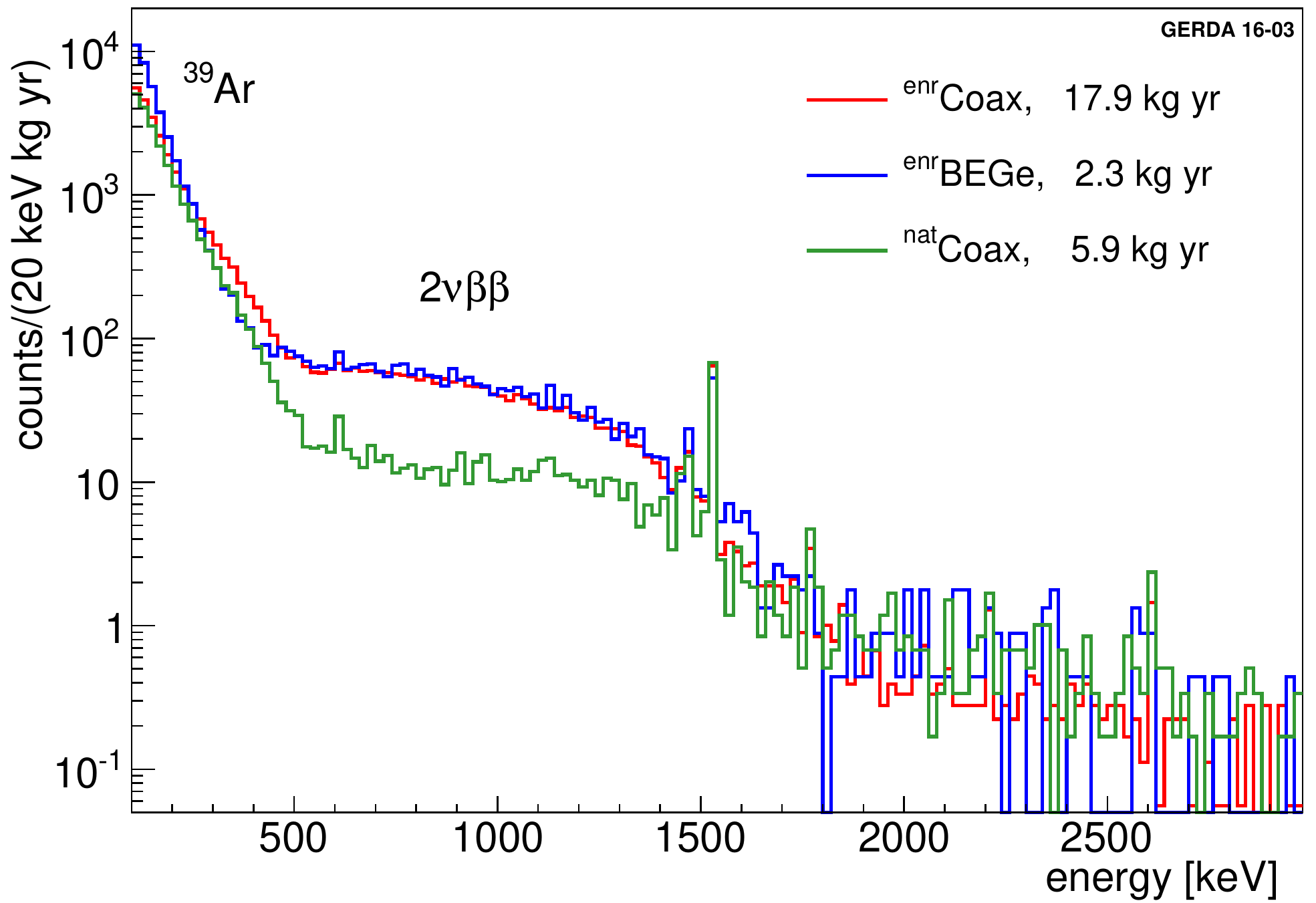}
\hspace*{2mm}
  \includegraphics[width=0.48\textwidth]{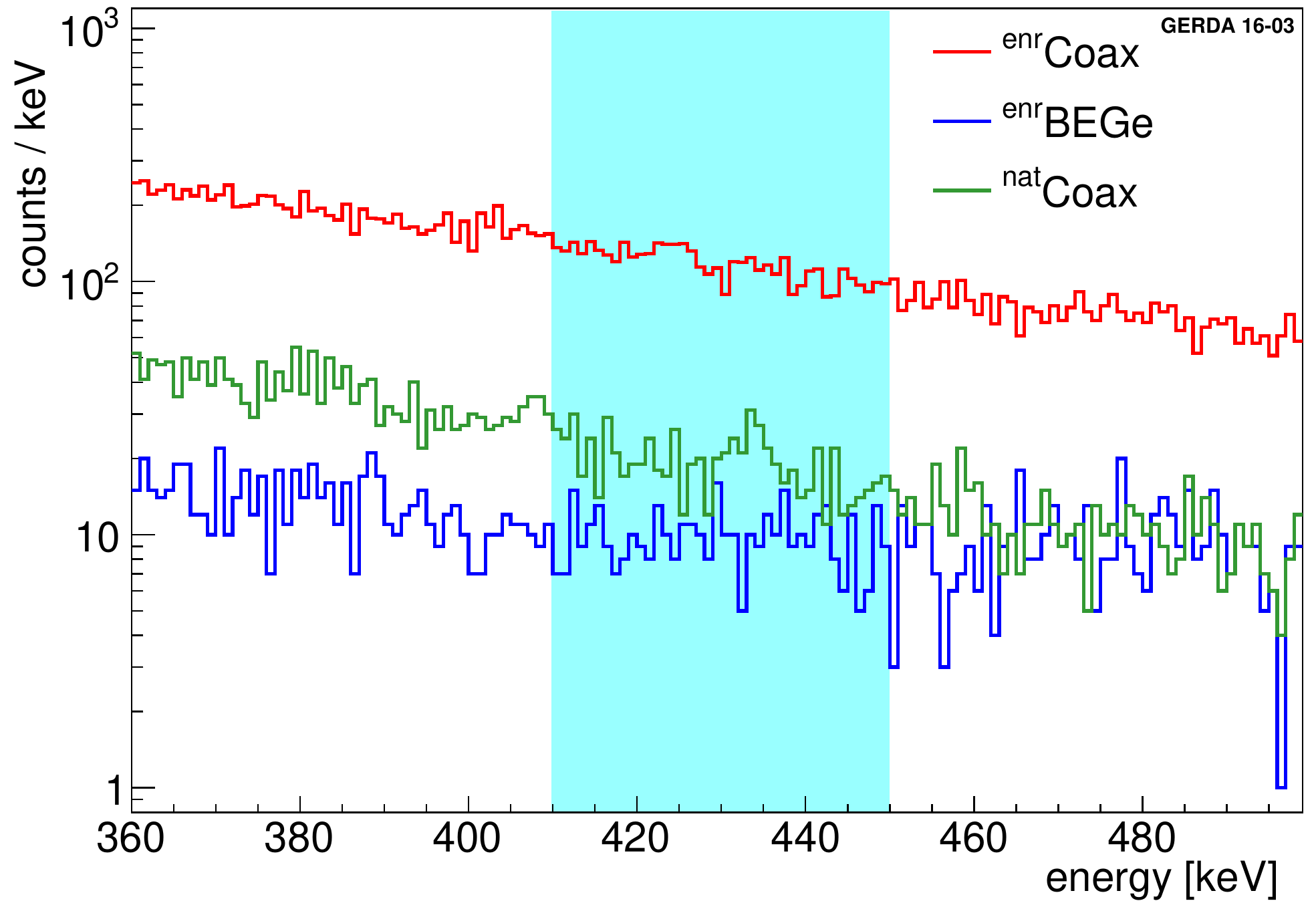}
\caption{\label{fig:3spectra}
      Energy spectra from the three data sets collected during
      \GERDA\ \phaseone. The left panel shows the energy spectra weighted with
      the product of life time and detector mass.  The right panel displays
      the energy region between 360 and 500~keV. The shaded area corresponds
      to the ROI defined between 410 and 450~keV
}
\end{center}
\end{figure*}
The data taking of \GERDA\ \phaseone\ started in November 2011 and ended in
May 2013. Until March 2012, the setup included 8 \geenr\ semi-coaxial and 3
\genat\ semi-coaxial detectors.  Two months later, two of the
\genat\ semi-coaxial detectors were replaced by five new
\geenr\ \bege\ detectors.  A higher background was observed during this period
(49\,d), therefore it was excluded from this analysis. The data taking was
separated into runs, with a duration of about one month each. Detectors which
showed instabilities during specific runs where removed from the analysis.
Two detectors showed instabilities from the very beginning of data taking.
Therefore, data collected from these detectors were discarded.  The total
collected data used for the search for \onecec\ of $^{36}$Ar correspond to a
life time of about 460\,d.  The data were divided into three different data
sets, one containing data from natural semi-coaxial detectors (labeled as
$^{\rm{nat}}$Coax), one containing data from enriched semi-coaxial detectors
($^{\rm{enr}}$Coax) and the last containing data collected by \bege\ detectors
($^{\rm{enr}}$BEGe).  The energy spectra from the three data sets are shown in
Fig.~\ref{fig:3spectra}. The left panel shows the energy spectra weighted with
the product of life time and detector mass.  The right panel displays the
energy region between 360 and 500~keV.  Indeed, in the region around
429.88~keV, enriched and natural detectors are characterized by different
contributions to the spectrum, in particular due to $2\nu\beta\beta$ decays
from \nuc{Ge}{76} in the enriched ones. In addition, \bege\ detectors are
considered as a separate data set because of the improved energy resolution
with respect to semi-coaxial detectors.  The main contribution to the spectrum
around 430~keV is due to $^{39}$Ar $\beta$ decays. The spectral shape is
different for \bege\ detectors due to the different detector geometry and
outer dead layer thickness.

Offline reconstruction of \GERDA\ data was performed within the
\gelatio\ software framework~\cite{gelatio:gelatio}.  Detector signals are
read out by charge sensitive preamplifiers and then digitized by 100~MHz flash
analog to digital converters (FADCs).  Preceded by a $\sim$80~\mus\ long
baseline, the charge signal rises up with a rise time of $\sim$1~\mus\ and is
followed by a $\sim$80~\mus\ long exponential tail.  The energy of each event
is estimated by applying an optimized Zero Area Cusp filter~\cite{zacfilter}
to the digitized signal.  Cuts based on the baseline slope, the number of
triggers and the position of the rising edge were applied to remove pile-up
events and accidental coincidences. All detected events within 8~\mus\ from
the muon veto trigger were also rejected.  Finally, an anti-coincidence cut
was applied to remove events with an energy deposition in more than one
detector.

The energy calibration was performed during dedicated calibration runs (every
one or two weeks) in which three $^{228}$Th sources were lowered to the
vicinity of the detectors.  In addition, the stability of the system was
continuously monitored by injecting test charge pulses into the input of the
preamplifiers. The energy dependence of the resolution was obtained for each
data set from the summed calibration spectra and then the value at the signal
peak position of 429.88~keV was derived.  The $^{42}$K background \gline\ at
1524.7~keV in the physics data was used to determine a correction factor in
case its energy resolution differed more than one standard deviation from the
one obtained during the calibrations.  To combine the different values into a
single value for the data set, the average of the energy resolution of each
detector was calculated weighted with the signal detection efficiency of the
detector.  The uncertainty on the resolution is primarily coming from the fit
of the resolution curve and is largest for the detectors that require the
correction factor~\cite{Giovannithesis}. The expected Full Width at Half
Maximum (FWHM) value at 429.88~keV is $4.08\pm0.20$\,keV for the
\nuc{Coax}{\rm nat}, $3.72\pm0.05$\,keV for the \nuc{Coax}{\rm enr} and
$2.01\pm0.10$\,keV for the \nuc{BEGe}{\rm enr} data set.  The systematic
uncertainty on the FWHM, estimated by comparing the resolution of the summed
calibration spectra to the average resolution of the single calibrations, is
$\pm$\,0.05~keV.

\section{Determination of the half-life of 0$\nu\rm{ECEC}$ of $^{36}$Ar}
A limit on the half-life $T_{1/2}$ of \onecec\ decay of $^{36}$Ar was
determined considering the data of \GERDA\ \phaseone\ discussed in
Section~\ref{sec:datataking}.  The region of interest (ROI) around the
signal, the 429.88~keV \ga\ line from the \onecec\ decay, is defined
between 410 and 450~keV.  The energy spectrum of coincidence events shows the
presence of the three \ga\ lines from $^\mathrm{108m}$Ag~\cite{excitedstates}.
$^\mathrm{108m}$Ag has a half-life of 418\,yr and undergoes electron capture into
the 6$^+$ excited state of \nuc{Pd}{108} with a probability of 91.3\,\%.  The
de-excitation of the daughter nucleus leads to three equally probable
\grays\ in the final state, with energies of 433.9\,keV, 614.3\,keV and
722.9\,keV.  The presence of $^\mathrm{108m}$Ag was also observed in the
screening measurements. For these reasons the 433.9\,keV \ga\ line from
$^\mathrm{108m}$Ag in the ROI was taken into account in the analysis.  The
determination of the detection efficiency and the analysis result are
discussed in the following.

\subsection*{\it Detection efficiency}
\begin{figure*}[th!]
\begin{center}
  \includegraphics[width=0.48\textwidth]{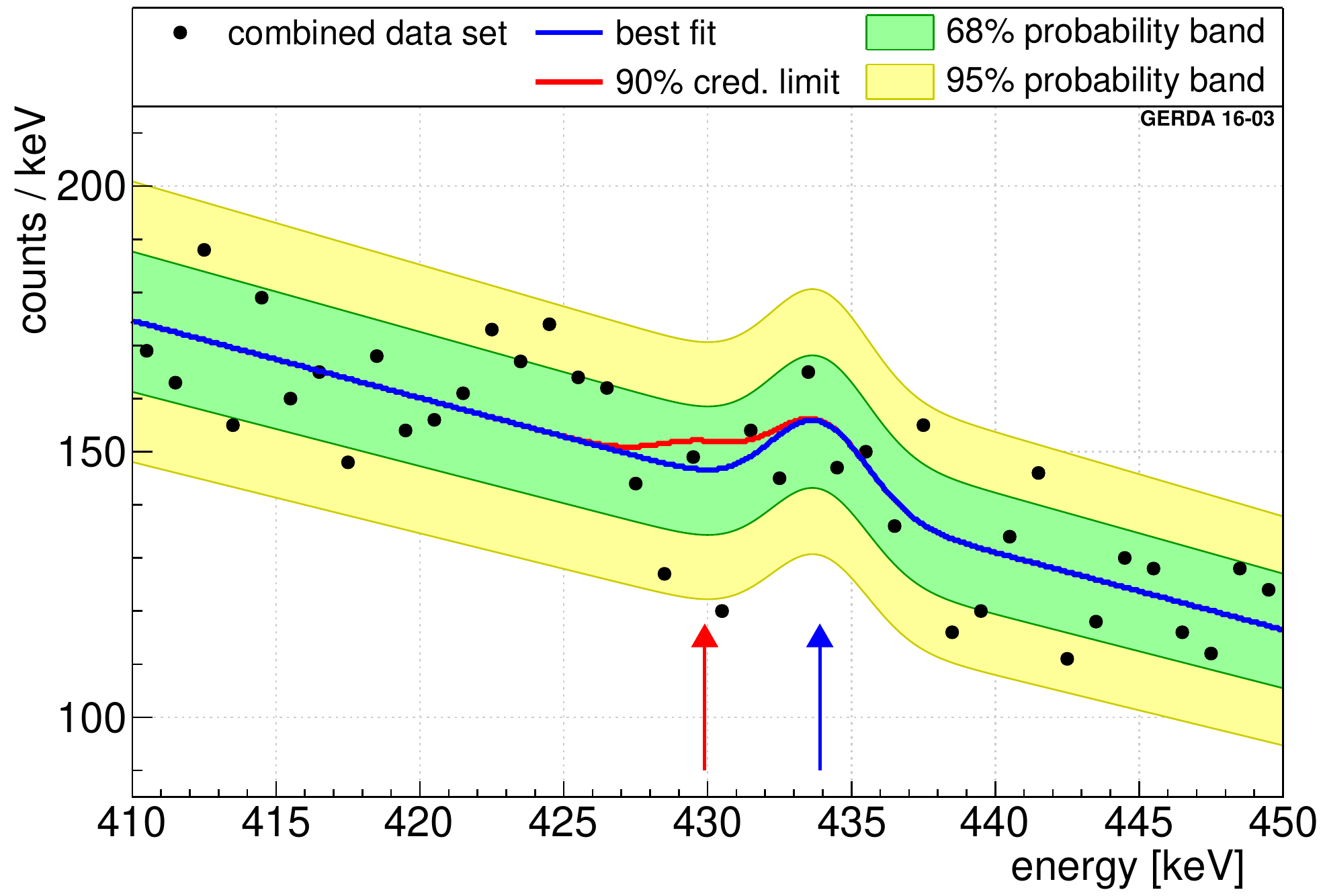} %_v3
\hspace*{2mm}
  \includegraphics[width=0.48\textwidth]{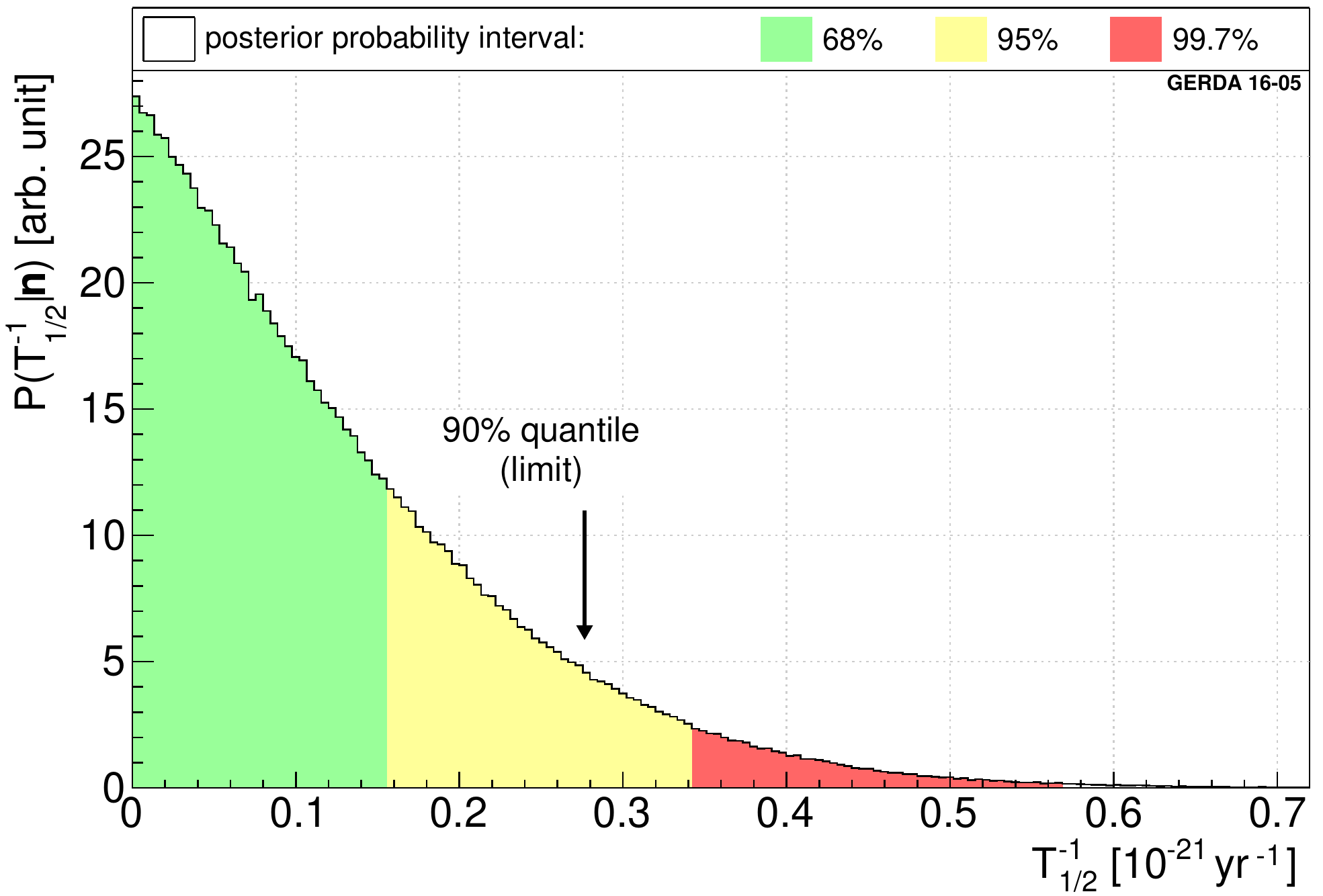}
\caption{\label{fig:bayesianfit}
     90\,\% C.I. Bayesian fit result for the inverse of the half-life on
     neutrinoless double electron capture of $^{36}$Ar. The left panel
     displays the experimental data from \GERDA\ \phaseone\ together with the
     best fit result (in blue) and the 90\,\% credibility interval limit (in
     red). The peak centered at 433.9~keV represents the best fit result for
     the \ga\ line from $^\mathrm{108m}$Ag. The arrows indicate the respective
     peak positions. The right panel shows the marginalized 
     posterior probability distribution for $T^{-1}_{1/2}$, where the arrow
     indicates the 90\,\% quantile from which the limit is derived
}
\end{center}
\end{figure*}
The detection efficiency $\varepsilon$ is defined as the number of
\grays\ which entirely deposit their energy inside a single
\GERDA\ detector. It has been determined by Monte Carlo simulations employing
the \mage\ software framework~\cite{mage} based on \geant~\cite{geant4}.
10$^{9}$ \grays\ with an energy of 429.88~keV were generated in a cylindrical
\lar\ volume with a radius of 67~cm and a height of 130~cm, centered around
the detector array. The considered volume corresponds to 1827~l of
\lar\ equivalent to 7.7~kg of $^{36}$Ar.  The contribution from
\grays\ originating from outside this volume to the number of full energy
depositions is less than the statistical uncertainty of 0.2\,\%.  The full
efficiency for each data set was derived by summing up the individual detector
efficiencies weighted for the life time of each run.  Their systematic
uncertainty is dominated by two main contributions: the uncertainty on the
Monte Carlo processes, whose effect on the efficiency was estimated to be
4\,\%, and the uncertainty on the dead layer of the germanium detectors. The
latter was estimated by independently varying for each detector the dead layer
values within $\pm$1 standard deviation.  This changes the efficiency of
8-10\,\% for a single semi-coaxial detector and 3.5-6\,\% for a single
\bege\ detector.  The uncertainty for the three data sets, calculated assuming
full correlation among the uncertainties of individual detectors, is 9.17\,\%
for the \nuc{Coax}{\rm nat} and \nuc{Coax}{\rm enr} data sets and 4.51\,\% for
the \nuc{BEGe}{\rm enr} data set.  The total systematic uncertainty on the
efficiency is obtained by summing in quadrature the two contributions and
amounts to 10\,\% for the \nuc{Coax}{\rm nat} data set, 10\,\% for the
\nuc{Coax}{\rm enr} data set and 6\,\% for the \nuc{BEGe}{\rm enr} data set.

Statistical uncertainties are negligible with respect to systematic ones.

\subsection*{\it Analysis}
The expected signal counts $S_d$ from neutrinoless double electron capture
from dataset $d$ are related to the half-life $T_{1/2}$ according to the
following relation
\begin{center}
\begin{equation}
S_d =\ln{2} \cdot \frac{\varepsilon_d}{T_{1/2}} \cdot \frac{
   N_A \cdot M_{LAr} \cdot f_{36} \cdot t}{m_{Ar}} \,\,\,,
\end{equation}
\end{center}
where $\varepsilon_d$ is the signal detection efficiency for data set $d$,
$N_A$ is the Avogadro constant, $t$ is the total life time, $M_{LAr}$ is the
mass of the \lar\ volume that was used for the efficiency determination,
$f_{36}$ the abundance of \nuc{Ar}{36} and $m_{Ar}$ the molar mass of argon.
\begin{table*}[t]
\begin{center}
\caption{\label{tab:summarydataset}
      Fit parameters values: FWHM is the Full Width at Half Maximum,
      $\varepsilon$ the signal detection efficiency, B$_{\rm{Ag}}$ the
      expected number of counts from the 433.9 keV $^\mathrm{108m}$Ag
      \ga\ line and B$_0$ the expected number of counts from the linear
      background component at the signal position
}
\footnotesize
\begin{tabular}{l|cccc}
\toprule
Data set & FWHM & $\varepsilon$ & B$_{Ag}$ &  B$_0$\\
              & (keV)   &                        & (counts) & (counts/keV)\\
\midrule
\nuc{Coax}{\rm nat} & 4.08 $\pm$ 0.20  & \baseT{(2.92\pm 0.29)}{-4} &
                                                                      41.9$^{+14.0}_{-12.9}$ & 18.3 $\pm$ 0.8\\[0.2cm]
\nuc{Coax}{\rm enr} & 3.72 $\pm$ 0.05	 & \baseT{(7.06\pm 0.71)}{-4} & 24.6$^{+18.6}_{-23.0}$ & 116.9 $\pm$ 1.8\\[0.2cm]
\nuc{BEGe}{\rm enr} & 2.01 $\pm$ 0.10   & \baseT{(1.11\pm 0.07)}{-4} &
                                                                       0.0$^{+5.3}$
                                                          & 9.7 $\pm$ 0.6\\
\bottomrule
\end{tabular}
\end{center}
\end{table*}
The unbinned likelihood function is defined as
\begin{eqnarray}
\mathcal{L} &=&
 \prod \limits_d \mu_d^{N_{d}} e^{-\mu_{d}} \prod \limits^{N_{d}}_{i} \frac{\lambda_{d,i}}{\mu_d}  \,\,\,,
\end{eqnarray}
where the product runs over all data sets $d$ and events $i$. $N_d$ is the total
number of events in the data set.  
$\lambda_{d,i} = \lambda_{d}(E_{d,i}|{\bf  p}_d)$ is the extended probability
density of finding an event with energy $E_{d,i}$ in dataset $d$ with a given
set of parameters ${\bf p}_d$. 
$\mu_{d}$ represents the total number of expected events in dataset $d$ over
the whole energy range $\mu_{d} = \int{\lambda_{d}(E|{\bf p}_d)dE}$.  In the
region of interest the background is in good approximation linear.
Therefore, $\lambda_{d,i}$ can be described as the sum of a linear background
contribution plus a peak from $^\mathrm{108m}$Ag and the signal peak from
\onecec\ of $^{36}$Ar
 \begin{eqnarray}
 \lambda_{d,i} &=&
  \frac{1}{\sqrt{2\pi}\sigma_{d}} 
  \left \{S_{d}\cdot
                   {\rm exp}\left[-\frac{(E_{d,i} - 429.88 + \delta_E)^2}{2\sigma_{d}^2}\right]
                   \right. \nonumber\\[2mm]
 &&\left.+ B_{\rm{Ag},d}\cdot
 {\rm exp}\left[-\frac{(E_{d,i} - 433.9 + \delta_E)^2}{2\sigma_{d}^2}\right]\right \} \nonumber\\[2mm]
 &&+ B_{0,d} + B_{1,d} \cdot (E_{d,i} - 429.88) \,\,\,, \label{eqn:expcounts}
 \end{eqnarray}
where $\sigma_{d}$ is the energy resolution ($\rm{FWHM}=2.35\cdot\sigma_{d}$),
$\delta_{E}$ a possible systematic shift in energy scale. $B_{0,d}$ and
$B_{1,d}$ describe the linear background and
$B_{\rm{Ag},d}$ the count expectation of the $^\mathrm{108m}$Ag peak.  A Bayesian
approach was used to extract the posterior probability density on
$T_{1/2}^{-1}$.  In total, the fit has 17 floating parameters, six describing
the signal peak ($\varepsilon_d$, $\sigma_{d}$), six for the linear background
($B_{0,d}$, $B_{1,d}$), three for the $^\mathrm{108m}$Ag peak ($B_{\rm{Ag},d}$).
$T_{1/2}^{-1}$ and $\delta_{E}$ are in common to all data sets.  The
parameters $\varepsilon_d$, $\sigma_{d}$ and $\delta_{E}$ are constrained by
Gaussian shaped prior distributions whose sigma is given by their systematic
uncertainty.  A flat prior is considered for the remaining parameters,
including the inverse half-life $T_{1/2}^{-1}$.  Furthermore, $B_{0,d}$,
$B_{Ag,d}$ and $T_{1/2}^{-1}$ are bound to positive values, while $B_{1,d}$ is
bound to negative values.  The best fit is defined as the mode of the global
posterior probability density and yields $T_{1/2}^{-1} = 0$, i.e. no signal
events from 0$\nu\rm{ECEC}$. The 90\,\% credibility limit of the half-life,
defined as the 90\,\% quantile of the marginalized posterior distribution, is
\begin{equation}
T_{1/2} > 3.6 \cdot 10^{21} \mbox{ yr (90\,\% C.I.).}
\end{equation}
The median sensitivity for the 90\,\% C.I. limit was estimated with toy Monte
Carlo simulations and is equal to 2.7 $\cdot$ 10$^{21}$~yr.  The sum spectrum
of all data sets around the ROI and the fit functions are displayed in
Fig.~\ref{fig:bayesianfit} together with the marginalized posterior
distribution for $T_{1/2}^{-1}$.

Systematic uncertainties are directly folded into the fit through the Gaussian
priors associated to parameters $\varepsilon_d$, $\sigma_{d}$ and
$\delta_{E}$.  They weaken the limit by about 0.3\,\%, which was evaluated by
fixing these 7 parameters and repeating the fit with the remaining 10
parameters. To test if the model described in Eq.~\ref{eqn:expcounts} is
sufficient, the p-value was calculated for the three data sets, as proposed in
Ref.~\cite{allenpvalue} using a 1 keV binning.  The obtained values are 0.96,
0.11 and 0.91 for the \nuc{Coax}{\rm nat}, \nuc{Coax}{\rm enr} and
\nuc{BEGe}{\rm enr} data sets respectively and indicate that the model
describes the data sufficiently well.  The fit result for $B_{Ag}$ shows the
presence of the 433.9 keV \ga\ line in the \nuc{Coax}{\rm nat} data set.  The
90\,\% C.I. limit is reduced by 10\,\% in case the presence of this line is
neglected in the fit.  The expectation value for the number of counts from the
$^\mathrm{108m}$Ag \ga\ line for the three data sets is reported in
Table~\ref{tab:summarydataset} together with the fit result for $B_0$ which
represents the number of events from the linear background component at the
signal peak energy of 429.88 keV (third term of Eq.~\ref{eqn:expcounts}).  In
the same table the efficiency values and the energy resolution are also
reported.

\section{Conclusions}
\gerda\ established the most stringent half-life limit on the radiative mode
of neutrinoless double electron capture of $^{36}$Ar with \phaseone\ data. The
limit is three orders of magnitude larger than previous results for this
isotope; however, it is still orders of magnitude far from the theoretical
prediction from QRPA calculations.

\begin{acknowledgements}
The \gerda\ experiment is supported financially by
   the German Federal Ministry for Education and Research (BMBF),
   the German Research Foundation (DFG) via the Excellence Cluster Universe,
   the Italian Istituto Nazionale di Fisica Nucleare (INFN),
   the Max Planck Society (MPG),
   the Polish National Science Centre (NCN),
   the Foundation for Polish Science (MPD programme),
   the Russian Foundation for Basic Research (RFBR), and
   the Swiss National Science Foundation (SNF).
 The institutions acknowledge also internal financial support.

The \gerda\ collaboration is grateful for useful discussions with V. Tretyak.
The \gerda\ collaboration thanks the director and the staff of LNGS for their
continuous strong support of the \gerda\ experiment.  Furthermore we
acknowledge the use of the CPU farm ATLAS of ZIH at TU Dresden for the Monte
Carlo simulations.
\end{acknowledgements}


\end{document}